# Multifractal Analysis and Local Hoelder Exponents Approach to Detecting Stock Markets Crashes

I. A. Agaev [1], Yu. A. Kuperin [2]

[1] *Division of Computational Physics, Saint-Petersburg State University*
*198504,Ulyanovskaya st., 1, Saint-Petersburg, Russia*
E - mail: `ilya-agaev@yandex.ru`

[2] *Laboratory of Complex System Theory, Saint-Petersburg State University*
*198504, Ulyanovskaya st., 1, Saint-Petersburg, Russia*
E - mail: `kuperin@jk1454.spb.edu`

**Abstract**

This paper is devoted to problem of detecting critical events at finiacial markets using methods of multifractal analysis. Namely, the local regularity of time-series is studied. As a result, one can find out a special behavior or signal of regularity before crashes. This spesial behaviour of local Hoelder exponents inherent in financial time series can be used in detecting critcal events or crashes at financial markets.

## 1. Introduction

According to academic point of veiw that markets are efficient, only the relevant negative information can cause a crush. This linear theory was firstly formulated in papers of Samuelson (1965) and Fama (1970, 1991). Analytically, the concept of information has been defined rigorously. Statistically, the notion of random walk has been generalized to Ito processes. But it seems that linear paradigm can not explain widespread critical events on financial markets. Furthermore there are some general ways of explaining market dynamics nowaday. Besides variety of papers, devoted to classical econometric standpoint like Cornell (1977), Frankel (1980) and Sharp et al. (2003), there are a lot of nonlinear techniques, looking for deterministic nonlinear and chaotic dynamics in different economical systems (Hilborn, 1994, Kantz, Schreiber, 1997, Bleaney et al., 1996) and in particular in the market. At last, a different recent fractal and multifractal techniques have been found to give a lot of useful information about financial time-series (Peters, 1991, Peters, 1994, Hilborn, 1994). Moreover, a multifractal model of assets returns has been developed (Mandelbrot et al., 1997).

Important step in studing of financial markets has been made in 1997 by Mantegna and Stanley, who have shown, that distribution of financial data fits to stable Levi distribution. Moreover, in the next papers (Stanley, 2001, Mantegna, 2000), studing cumulative distribution for stocks, the authors concluded that stocks have power law behavior on tails of distribution. This was an evidence of presence of scaling in investigated data. This important step gave rise to applying the methodology of multifractal theory to various financail instruments. In our work for numerical study we used the *Fraclab* – a Matlab toolbox based on fractal and multifractal approaches (http://fractales.inria.fr/).

## 2. Local Hoelder Exponents

Before discussing an application of local Hoelder exponents technique to study of financial data, let us remind some useful definitions.

**Definition 1**: *Function $f(x)$ belongs to class of Hoelder functions of order $\alpha$, if*

$$|f(t+h) - f(t)| < const \cdot h^\alpha; \ t, h \in \mathbb{R}; 0 \leq \alpha \leq 1 \quad (1)$$

*In case then $\alpha$ depends on t, i.e. $\alpha \to \alpha(t)$, then $\alpha(t)$ called a local Hoelder exponent.*

Extreme cases of Hoelder functions are well-known. As $\alpha = 1$ it coincide with class of differentiable functions. As $\alpha = 0$ it coincide with class of function with nonremovable discontinuity. Thus, the class of Hoelder functions is intermediate between smooth and discontinuous functions and Hoelder exponent is the measure of function irregularity.

It is something common in modern econophysics to suppose, that financial time-series can be represented by local Hoelder functions. Obviously, periods of critical events cause significant and sudden changes in corresponding time series. Hence, the calculation of the local regularity of time-series should give a new power tool for detecting critical events.

### 2.1 Methodology of multifractal analysis

The main question of multifractal analysis of time series, in general, and obtaining of local Hoelder exponents, in particularly, is how to get them from 1D-data. The usual way is following:

*Let $Y(t)$ is an asset price, then the measure associated with time series in question can be defined for example like:*

$$X(t, \Delta t) = (\ln Y(t + \Delta t) - \ln Y(t))^2 \quad (2)$$

*Then one should divide the time interval [0,T] into N intervals of length $\Delta t$ and define sample sum:*

$$Z_q(T, \Delta t) = \sum_{i=0}^{N-1} |X(i \cdot \Delta t, \Delta t)|^q \quad (3)$$

*In terms of sample sum one can define the scaling function:*

$$\tau(q) = \lim_{\Delta t \to 0} \frac{\ln Z_q(T, \Delta t)}{\ln \Delta t} \quad (4)$$

*The spectrum of fractal dimensions of squared log-returns X(t,1) is defined as:*

$$D_q = \frac{\tau(q)}{q-1} \quad (5)$$

*Finally to obtain local Hoelder exponent one should make one more step – differentiation of τ(q):*

$$\alpha = \frac{d\tau(q)}{dq} \quad (6)$$

Some remarks should be done here. First of all, one now has a possibility to determine is the given time-series multifractal or it is not. Namely, if $D_q \neq D_0$ for some real $q$, then $X(t,1)$ is multifractal. For monofractal time series scaling function $\tau(q)$ is just linear.

Multifractal time series can be characterized now by local Hoelder exponents $\alpha(t)$: $X(t,\Delta t) \sim (\Delta t)^{\alpha(t)}$. In the classical asset pricing model (geometrical Brownian motion) $\alpha(t) = 1$.

## 3. Results

In present paper two different types of financial markets have been studied. The first one is the USA stock market which is treated now as the most developed and strong market over the world. And it is the most "efficient" in academic sense. The second one is the Russian stock market and associated with them the Russian currency market. Both of them are an exact antithesis to the USA stock market. They are developing small markets which should be far from efficiency. In the present study we have shown that the proposed methodology of the local Hoelder exponents is applicable and works well in USA stock market as well as in the Russian currency market.

### 3.1 USA Stock Market (DJIA index)

Let us remind the most significant crisis at USA stock market from January 1980 until January 2003. All of them had different reasons and effects on the US economy. According to (Sornette D., 2003) they are:

- **Summer 1982.** Sometimes it called an oil crisis due to its main reasons – an oil embargo of arabian countries, high prices on oil, high inflation processes and declined by South America countries debt pays. On the stock market at that time a lot of sudden crashes were observed.

- **October 1987.** The crash of October 1987 and its Black Monday on October 19 remains one of the most striking drops ever seen in stock markets, both by its overwhelming amplitude and its sweep over most markets worldwide. On Black Monday, the Dow Jones industrial average lost 508 point, or 22.6% of its value. There are a lot of reasons, but experts emphasize three of them: imperfect computer trading system, trade and budget deficits and high overvaluation of stock market.

- **Autumn 1998.** The worldwide known Asian Crisis came to USA in 1998. It had grave consequences on USA economy because internationality of USA corporations. Most of them had an productive capacity in crisis region. Also strong overvaluation of USA stock market should be mentioned.

- **September 2001.** In September 2001, USA economy had a lot of problems with budget deficit and high-tech crisis. And terrorist attack on World Trade Center become a catalyst of drop on USA Stock markets.

In Figure 1 one can see a picture with log-price of DJIA (1980-1988) at the top panel and calculated values of the local Hoelder exponents at the bottom panel. Both time series are at the same time-scale, hence, their values corrspond each other.

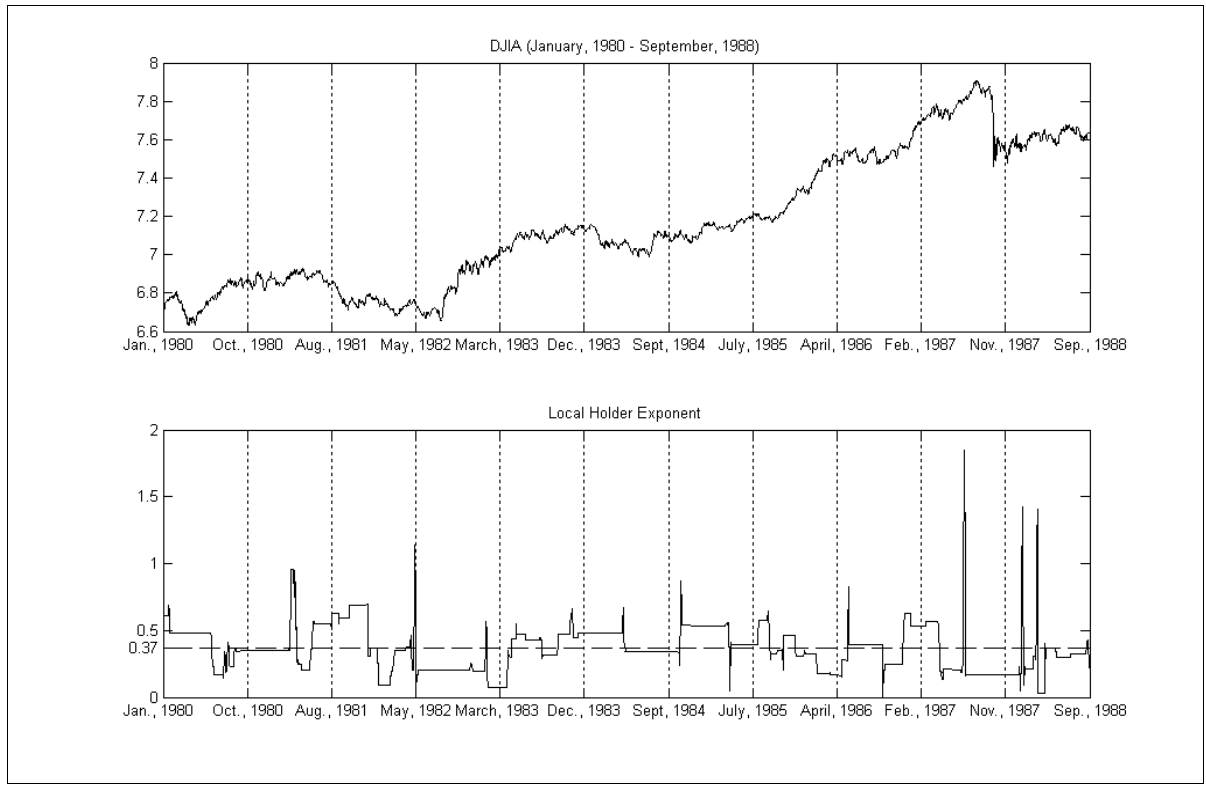

*Figure 1. Log-prices of DJIA and corresponding local Hoelder exponents (1980-1988) Dashed line represents the mean value of the Hoelder exponents..*

One should remark, that in Figure 1 there are four high values of local Hoelder exponents. They correspond to May 1982, June 1987 and to beginnig of 1988. Remind that the Hoelder exponent between 0 and 1 means that the signal is continuous but not differentiable at the considered point. In addition, the lower value of the exponent means the more irregularity in the signal. Looking at the original signal, it appears obvious that the log-prices is almost nowhere smooth, which is consistent with the values of the local Hoelder exponents.

Now let us note, that important events in the log-prices of DJIA have a specific signature in values of the local Hoelder exponents: periods where critical events took place are characterized by sudden increase in regularity, which passes above 1, followed by very small values of regularity for long period of time. It is reasonable to call such specific behaviour of local Hoelder exponents as *crash pattern*.

Let us consider some examples. Only two peaks in the Figure 1 followed by wide area of small values (which are well below the mean ) of local Hoelder exponents, i.e. they are *crash patterns*, and both of them corresponds to crisis described earlier.These are crisis of summer 1982 and October 1987.

Consider now another time interval (see Figure 2) which contains many points with low values of local Hoelder exponents and with a few isolated very regular points (with exponent larger than 1). There are six areas with a few very regular points (autumn 1995, autumn 1998, spring 2000, summer 2001, winter 2002, winter 2003). But only two of them are the *crash* *patterns*, i.e. they have the second feature – plain area with points of low regularity. And both them are correspond to crisis of 1998 and 2001 years.

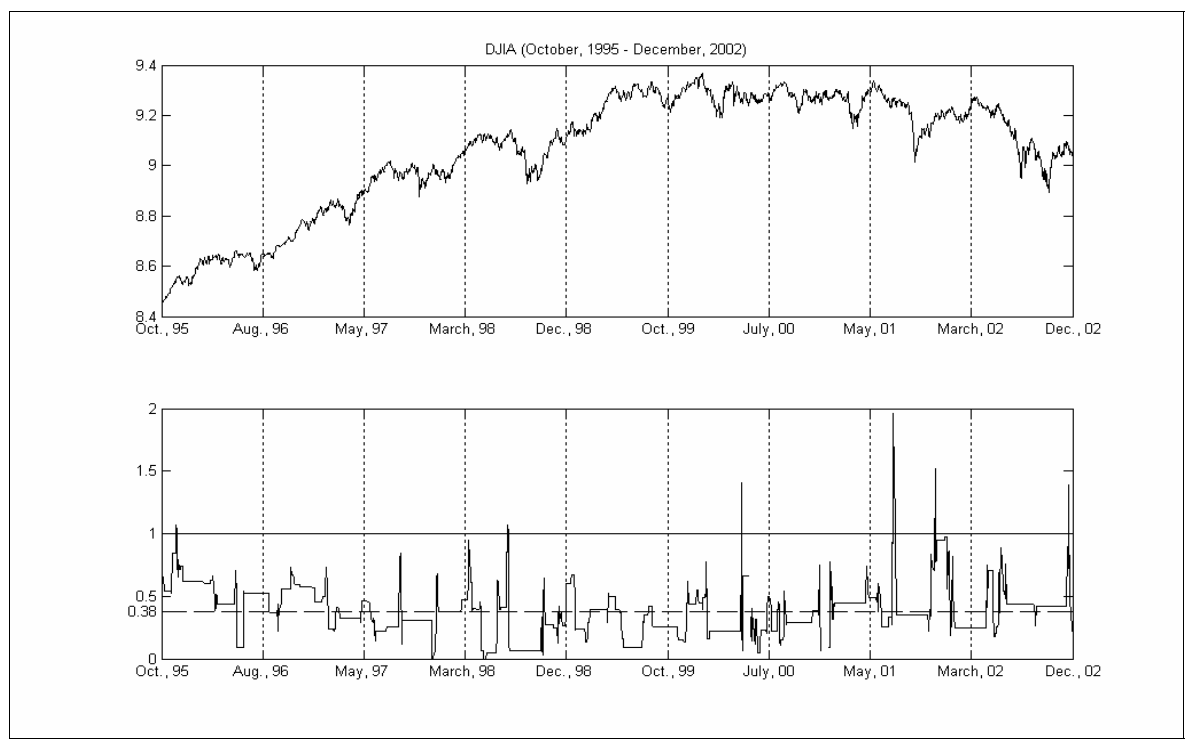

*Figure 2. Log-prices of DJIA and corresponding local Hoelder exponents (1995-2002). Dashed line represents the mean value of the Hoelder exponents. The solid line represents a level of differentiability.*

In spite of the approach we used is very simple and empirical in fact, it shows that changes with high amplitude in this particular financial time-series generally correspond to regions with very low values of the local regularity, with the presence of high regular points. Such a behavior can be observed in other financial time series.

Below we consider Russian currency market, which has a very special dynamics and strongly differs from almost "effective" USA stock market.

## 3.2  Russian Currency Market (RUR/USD Exchange Rate)

In this Section we restrict ourselves the exposition of the obtained results and their discussions. The choice of the time period analysed as well as the establishing of the reason of crashes on the Russian currency market will be published elsewhere. We have to mention only that chosen periods, i.e. 1992 – 1993 and 1994 - 1999 contain the most interesting and strong crashes to be analysed.

Figure 3 shows the September 1993 crisis on Russian currency market. As before, one can observe the special "signal" of strong perturbation. There is only one area, which has both indicators of perturbations, i.e. *crash pattern*. This is the point with very high value of regulation, followed by plain area of low-regular points. From the visual inspection of graphs in Figure 3 one can conclude that even at "inefficient" market in question the methodology of local Hoelder exponents works correctly.

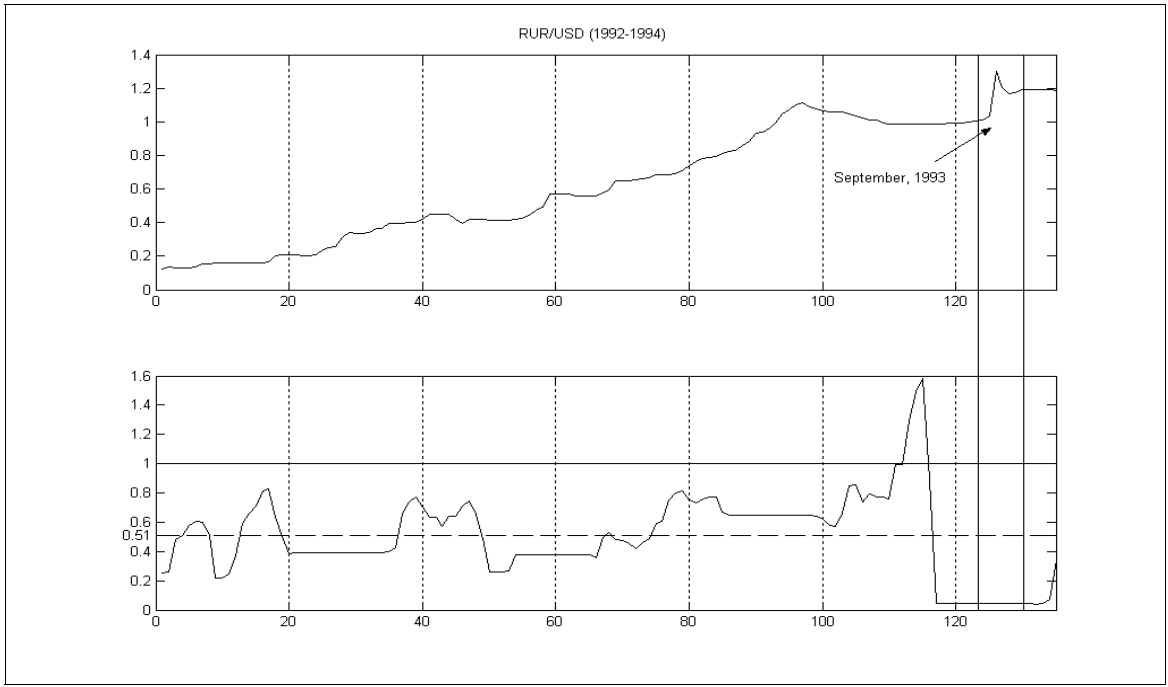

*Figure 3. RUR/USD exchange rate and corresponding local Hoelder exponents (1992-1993). Vertical lines locate the time perid of crsis. Dashed line represents the mean value of the Hoelder exponents. The horizontal solid line represents a level of differentiability*

Figure 4, which corresponds to another period, namely 1994-1999, again shows excellent results in the problem of crisis detecting.

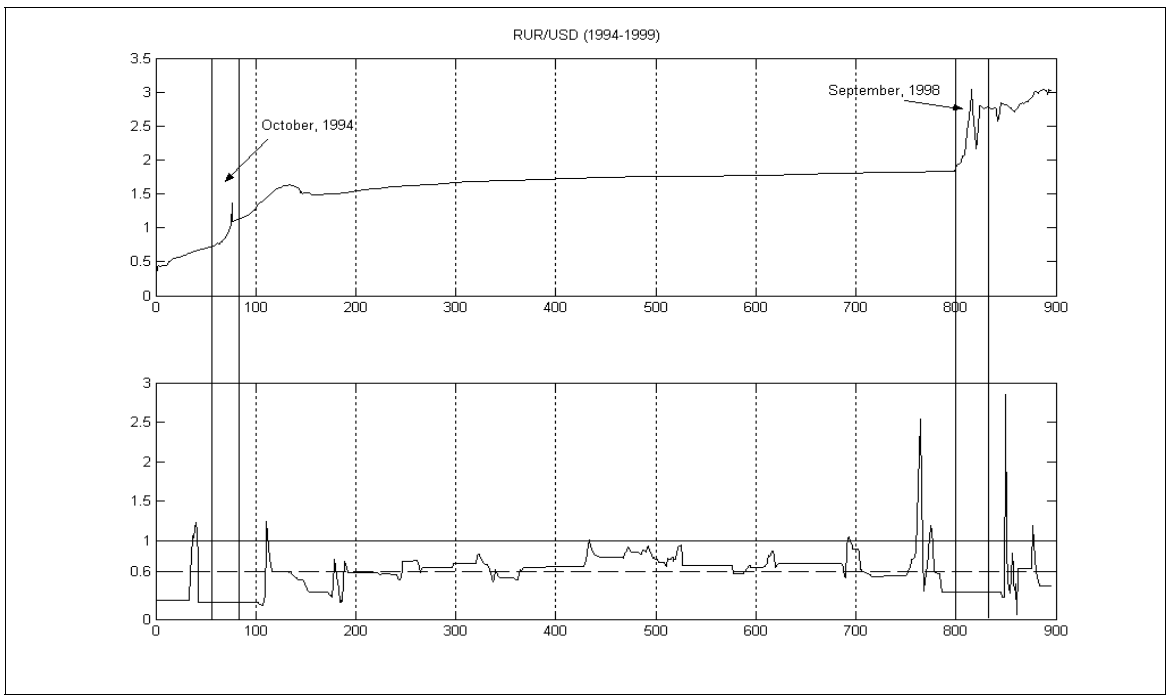

*Figure 4. RUR/USD exchange rate and corresponding local Hoelder exponents (1994-1999). Vertical lines locate the time perid of crsis. Dashed line represents the mean value of the Hoelder exponents. The horizontal solid line represents a level of differentiability*

# 4. Conclusions

In the present study we have shown that suggested method of local Hoelder exponents or extracting the *crash patterns* for analysis of the critical events on the financial markets and for detecting crashes on them works stable enough. Moreover the method is applicable to markets with essentially different dynamics. We have demonstrated this fact using the data characterised USA stock market and Russian currency market. It is important to note that the approach we used can be applied not only to a posteriori detecting of market crashes but also for a priori prediction of them. The power peak in the local Hoelder spectrum with the next very low plateau (*crash pattern*) can be evidently treated as a precursor of oncoming critical event on the market. In order to verify this it is enough to cut off the analysed time series directly before crisis and recalculate local Hoelder exponents. Even for truncated time series the *crash pattern* will appear. It means that the appearance of *crash patterns* can be used in on-line monitoring of the market.